\documentclass[conference]{IEEEtran}
\IEEEoverridecommandlockouts
\usepackage{cite}
\usepackage{amsmath,amssymb,amsfonts}
\usepackage{algorithmic}
\usepackage{graphicx}
\usepackage{textcomp}
\usepackage{xcolor}
\def\BibTeX{{\rm B\kern-.05em{\sc i\kern-.025em b}\kern-.08em
    T\kern-.1667em\lower.7ex\hbox{E}\kern-.125emX}}
\begin{document}

\title{Computational Electromagnetics with the RBF-FD Method\\
{\footnotesize
\thanks{The authors would like to acknowledge the financial support of the ARRS in the framework of the research core funding No. P2-0095, the Young Researcher program PR-12347 and research project J2-3048.}
}}

\author{
	\IEEEauthorblockN{Andrej Kolar-Požun}
	\IEEEauthorblockA{\textit{Parallel and Distributed Systems Laboratory,} \\
\textit{'Jožef Stefan' Institute,}\\
Ljubljana, Slovenia \\
and \\
\textit{Faculty of Mathematics and Physics,} \\
\textit{University of Ljubljana,} \\
Ljubljana, Slovenia \\
andrej.pozun@ijs.si}
\and
\IEEEauthorblockN{Gregor Kosec}
	\IEEEauthorblockA{\textit{Parallel and Distributed Systems Laboratory,} \\
\textit{'Jožef Stefan' Institute,}\\
Ljubljana, Slovenia \\
gregor.kosec@ijs.si}
}
\maketitle

\begin{abstract}
One of the most popular methods employed in computational electromagnetics is the Finite Difference Time Domain (FDTD) method. We generalise it to a meshless setting using the Radial Basis Function generated Finite Difference (RBF-FD) method and investigate its properties on a simple test problem.
\end{abstract}

\begin{IEEEkeywords}
meshless, rbf-fd, electromagnetics, fdtd, yee, maxwell
\end{IEEEkeywords}

\section{Introduction}
Accurate simulations of electromagnetic field propagation are of great importance in further development of wireless networks.
One of the most popular simulation procedures is solving the Maxwell's equations using the Finite Difference Time Domain (FDTD) method, which was proposed by Kane S. Yee over 50 years ago \cite{Yee}.
While Yee's algorithm provides accurate and reliable results, the fact that it is based on a grid-like discretisation is its weakness as it becomes impractical to describe fine geometric features, which is necessary for many real-world applications such as wave propagation in irregular domains or electromagnetic scattering on antennas of unusual shapes.

Several alternative procedures have been developed that remedy this \cite{soFar1,soFar3,soFar5, soFarMeshless1, soFarMeshless3}. Among these are meshless methods, which are especially suitable for handling irregular domain boundaries, as they can work on scattered nodes making it easy to discretise complex geometries \cite{slak}. While there have already been some successful attempts at modelling electromagnetic phenomena in the scope of meshless methods \cite{soFarMeshless1,soFarMeshless3}, none have really gained widespread use.

Despite that, we take yet another meshless approach to model the electromagnetic fields. In particular, using the Radial Basis Function Generated Finite Difference (RBF-FD) method we generalise the FDTD method. After introducing our approach we investigate some of the obstacles encountered in our further simulations.

More precisely, the paper is structured as follows:
The following section briefly lists the methods used including the original Yee's algorithm and the RBF-FD method.
Afterwards, our test problem is presented in Section 3, while the results are shown and discussed in Section 4.

\section{Methods}
Time evolution of electromagnetic fields is governed by the Maxwell's equations
\begin{align}
\mu_0 \frac{\partial H_x}{\partial t} =& - \frac{\partial E_z}{\partial y}, \\ 
\mu_0 \frac{\partial H_y}{\partial t} =&  \frac{\partial E_z}{\partial x}, \\
\varepsilon_0 \frac{\partial E_z}{\partial t} =& \frac{\partial H_y}{\partial x} - \frac{\partial H_x}{\partial y},
\end{align}
where we have limited ourselves to two dimensions and the TMz (Transverse Magnetic to the z direction) propagation mode meaning that only the $H_x, H_y, E_z$ fields and their $(x,y)$ dependence need to be considered.
$\mu_0$ and $\varepsilon_0$ are the vacuum permeability and permittivity respectively.
 
\subsection{The FDTD algorithm}
As hinted already by its name, the FDTD algorithm is in essence just a finite difference method - the partial derivatives appearing in Maxwell's equations are approximated with central differences, transforming them into a linear system.

The component that sets the algorithm apart from the usual finite difference methods is a particular way in which the space and time are discretised - the Yee grid.
We define, for indices $(n, i, j)$
\begin{align}
E_z^{n,i,j} &= E_z(t = n \Delta t,x = i \Delta s,y = j \Delta s),\\
H_x^{n,i,j} &= H_x((n-0.5) \Delta t, i \Delta s,(j+0.5) \Delta s), \\
H_y^{n,i,j} &= H_y((n-0.5) \Delta t, (i+0.5) \Delta s, j \Delta s),
\end{align}
where $\Delta t$ and $\Delta s$ are the chosen spacings in time and space direction, respectively.

This allows us to discretise the Maxwell's equations and get the explicit FDTD update equations:
\begin{align}
H_x^{n+1,i,j} = H_x^{n,i,j} -\frac{\Delta t}{\mu_0 \Delta s} &(E_z^{n,i,j+1}-E_z^{n,i,j}), \\
H_y^{n+1,i,j} = H_y^{n,i,j} + \frac{\Delta t}{\mu_0 \Delta s} &(E_z^{n,i+1,j} - E_z^{n,i,j}), \\
E_z^{n+1,i,j} = E_z^{n,i,j} + \frac{\Delta t}{\varepsilon_0 \Delta s} &(H_y^{n+1,i,j}-H_y^{n+1,i-1,j}  \nonumber \\
- &(H_x^{n+1,i,j}-H_x^{n+1,i,j-1})).
\end{align}
The staggering of the Yee grid causes the above method to be of second order in both space and time. A more detailed introduction to the FDTD method along with the necessary prerequisite knowledge of finite differences and electromagnetics can be found in \cite{EMPBook}.

\subsection{The RBF-FD Method}
The RBF-FD method allows us to obtain an approximation for a given linear differential operator $\mathcal{L}$ using only the positions of the discretisation nodes $(x_i, y_i)_{i=1}^N$ without any additional information about them.
This is done by associating to each node $(x_i,y_i)$ its stencil $S_i$ - a set of nodes used for the approximation, commonly taken to be its $ss$ (stencil size) nearest neighbours. The operator acting on a function is then expressed as a linear combination of function values in the stencil as
\begin{equation}
\mathcal{L} f(x_i, y_i) \approx \sum_{j \in S_i} w_j f(x_j, y_j),
\end{equation}
where the weights $w_j$ can be determined from the locations of the nodes alone. The details exceed the scope of this paper and we refer the reader to chapter 5 of \cite{RBFFD}.

\subsection{Our Meshless FDTD Generalisation}
We now describe our attempt at generalising the FDTD method to a meshless setting. The time coordinate has a simple geometry so we keep this part of the FDTD and stagger the fields in time. The difference comes with how the spatial derivatives are treated. There isn't a clear way on how the spatial staggering should be done if the nodes are allowed to be positioned artbitrary so we have decided to abandon it and calculate both the $E$ and $H$ fields at the same space points. This results in the following update equations
\begin{align}
H_x^{n+1}(x,y) = H_x^n (x,y) - \frac{\Delta t}{\mu_0} &\mathcal{L}_{y} (E_z^n)(x,y), \\
H_y^{n+1}(x,y) = H_y^n (x,y) + \frac{\Delta t}{\mu_0} &\mathcal{L}_{x} (E_z^n)(x,y), \\
E_z^{n+1}(x,y) = E_z^n (x,y) + \frac{\Delta t}{\varepsilon_0} (&\mathcal{L}_{x} (H_y^{n+1})(x,y) \nonumber \\- &\mathcal{L}_y (H_x^{n+1})(x,y)),
\end{align}
where $\mathcal{L}_x$ is a RBF-FD approximation of $\partial_x$ and analogously for $\mathcal{L}_y$. RBF-FD requires us to make a choice of a Radial Basis Function (RBF) to be used in the approximation. We employ the Polyharmonic Splines (PHS) of degree $3$ as our RBF of choice, as they possess no shape parameter, eliminating the additional work that would come with determining its suitable value. Additionally, our approximation is augmented by monomials up to degree $m=2$ inclusive, which causes the derivatives to be second order accurate ~\cite{Bayona}, as is the case in the FDTD. We start with a minimal stencil size of $ss=6$, corresponding to a pure monomial approximation ~\cite{Bayona}.

\section{Problem setup}
The considered problem is propagation of the electromagnetic field from a single source in an otherwise empty medium (vacuum).
The domain is a square $1 \leq i,j \leq 600$, where the discretisation distance is $\Delta s = 1$. The time step is given by the Courant limit $\Delta t = \frac{S_c \Delta s}{c_0}$, where $c_0$ is the speed of light in the vacuum and $S_c$ is the Courant number, taken to be $\frac{1}{\sqrt{2}}$ to minimise dispersion \cite{EMPBook}.

Near the middle of the domain, we have a sinusoidal source: $E_z^{n,300,300} = \sin \left(2 \pi f n\right)$ with a frequency of $f = S_c/30$. This is a hard source, meaning we manually set the field to this value after each time step. The simulation will stop at $n=300$, which is before the waves reach the boundary so we do not need to take any boundary effects into account.

We will compare both the FDTD and our described method on this setup. While meshless methods are usually not meant to be used on a grid, this is only to test whether our RBF-FD formulation can replicate the correct physics. We don't expect more accurate results than those obtained by FDTD - the advantage of our method would come from being able to operate on scattered nodes.

\section{Results}
\begin{figure}
    \centering
    \includegraphics[width=\linewidth]{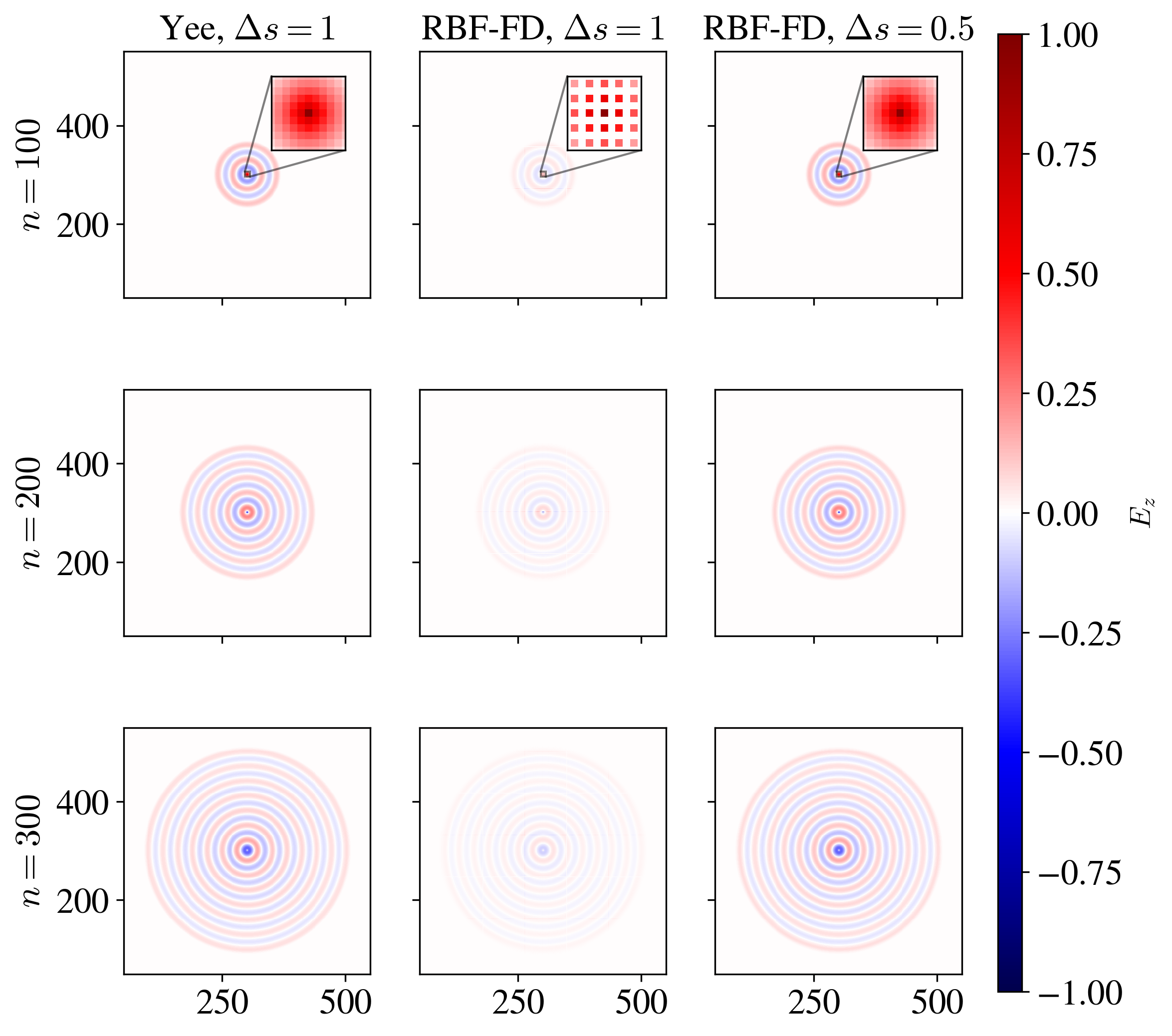}
    \caption{Solution snapshots for both the FDTD and RBF-FD methods. In the $\Delta s = 0.5$ case we have hidden the irrelevant nodes. A zoomed-in portion of the snapshot is shown for the $n=100$ case.}
    \label{fig:contours}
\end{figure}

Figure \ref{fig:contours} shows the resulting simulations. A problem can immediately be spotted: certain grid points are not updated properly and remain at their initial value of zero, resulting in a checkerboard-like pattern.

\begin{figure}
    \centering
    \includegraphics[width=\linewidth]{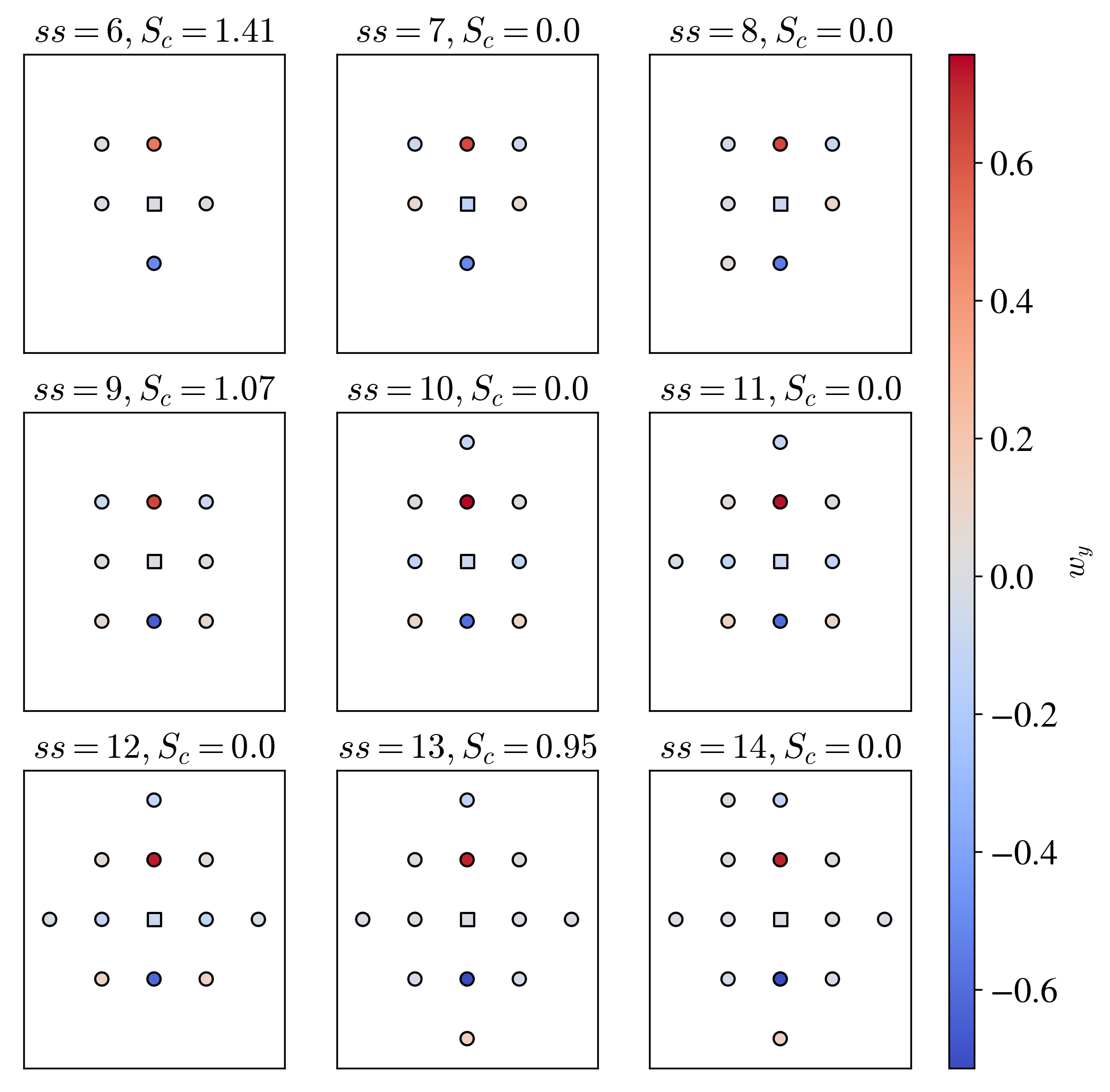}
    \caption{Different stencils and their corresponding stability limit Courant numbers. The center node is marked with a square and the node colours correspond to the RBF-FD weights $w_y$ for $\partial_y$.}
    \label{fig:stencilsStability}
\end{figure}

The reason lies in the approximation of the spatial derivatives appearing in Maxwell's equations. As seen in Figure \ref{fig:stencilsStability}, RBF-FD with a stencil size of $ss=6$ replicates the ordinary central difference formulas for the first derivative. These central differences are such that the derivative at given point $(x_i,y_i)$ depends only on the function values at the neighbouring nodes, not on the node $(x_i,y_i)$ itself. Plugging this into the update equations for our case, it turns out that the values of $E_z$ at  nodes, that are an odd number of grid points away from the source in either the $x$ or $y$ direction, are never updated. A similar pattern arises for fields $H_x, H_y$.

If we consider only the fields that are getting updated (i.e. ignoring the update equations for the irrelevant ones), we actually get a Yee grid with a discretisation distance of $\Delta s = 2$. This motivates us to repeat the simulation with $\Delta s =0.5$ (while keeping the Courant number fixed) and only plot the integer coordinates, hiding the nodes, which we know are not updated properly. The resulting solution snapshots look much closer to the ones given by the FDTD simulation. We have also numerically confirmed that we do indeed get the same quantitative results as with the usual FDTD method. Deviating from this setup in some way, such as by increasing the stencil size, will therefore allow us to obtain generalised versions of the FDTD.

Using a finer discretisation only to then throw out a big portion of the nodes is computationally wasteful. One attempt to migitate this issue is to use non-centered stencils, as then the algorithm will not reproduce the usual central difference formulas. However, in this case a new problem arises - stability. It is well known that in the FDTD method we require our time step to be sufficiently small for a stable time evolution. A similar limitation also holds for other choices of the solution procedures with different stability conditions.

In order to assess the stability of our method, we have performed the Von Neumann stability analysis described in \cite{EMPBook}. A plane wave ansatz $E^n_z (x,y) = E^n_{z0} e^{i (k_x x+k_y y)}$ (and similarly for $H_x, H_y$)  is plugged into our update equations to obtain a relation
\begin{equation}
\begin{pmatrix} H^{n+1}_{x0} \\ H^{n+1}_{y0} \\ E^{n+1}_{z0} 
\end{pmatrix} = A
\begin{pmatrix} H^{n}_{x0} \\ H^{n}_{y0} \\ E^{n}_{z0} 
\end{pmatrix}.
\end{equation}
We then check at which time step $\Delta t$ or, equivalently, at which Courant number $S_c$, the spectral radius of the matrix $A$ (the magnitude of its largest eigenvalue) exceeds $1$. This gives sufficient conditions for stability. 

The results can be seen in Figure \ref{fig:stencilsStability}, which shows that the considered asymmetric stencils are unstable. An exception is the case of $ss=6$, where the node that is ruining the symmetry has a zero weight effectively making the stencil symmetric. Note that this is the previously mentioned case of RBF-FD reproducing the central difference formulas. 

Overcoming the stability issue is not at all trivial and will require a lot of care when attempting to use the RBF-FD method for different stencils, especially in the case of scattered nodes, where many possible stencil shapes can arise of varying regularity. This is a whole topic of its own and we have not delved into it any further at this time. Instead, we look at the behaviour of the method for stencil sizes above the minimal $ss=6$ case that are still stable: $ss=9,13,25$\footnote{while the $ss=25$ case is not covered in Figure \ref{fig:stencilsStability}, it is a symmetric stencil and turns out to be stable.}.
 
\begin{figure}
    \centering
    \includegraphics[width=\linewidth]{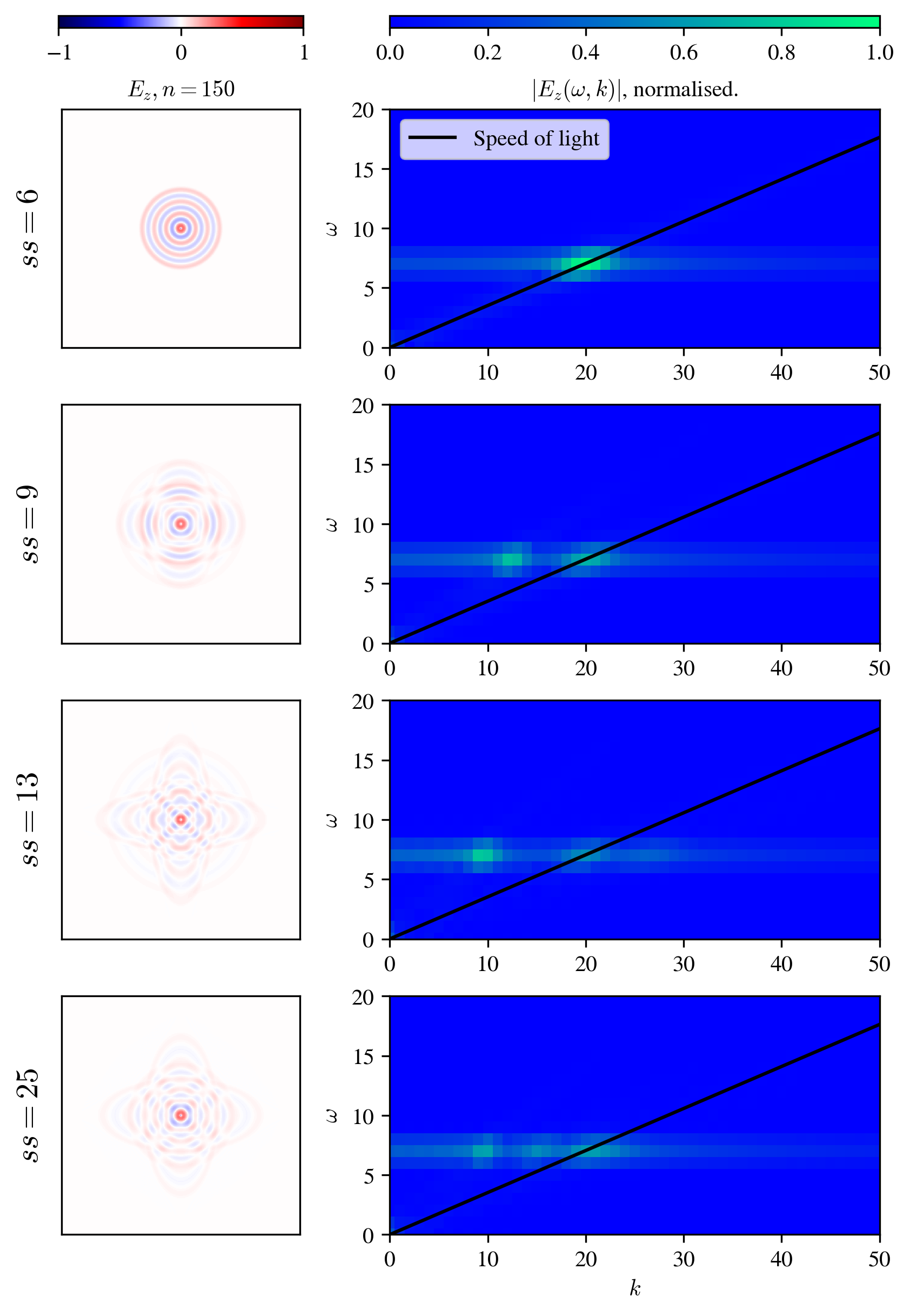}
    \caption{Solution snapshots at $n=150$ and the normalised Fourier component magnitudes, plotted for various stencil sizes. $\omega$ and $k$ are the temporal and spatial frequency indices respectively. The speed of light is shown with a black line.}
    \label{fig:dispersion}
\end{figure}

These stencils are centered and turn out to also exhibit checkerboard-like patterns. So for the purpose of visualisation we again set $\Delta s = 0.5$ and only display the integer coordinates. 
We continue to simulate at $S_c = \frac{1}{\sqrt{2}}$, but we have checked that the following observations occur also for different time steps.
Figure \ref{fig:dispersion} shows the results and highlights another important issue - dispersion. The simulation fails to produce the correct physics as some superluminal modes start propagating.  
This can be concisely seen in the Fourier space: Considering the solution slice $E_z(n,i,j=300)$ and performing the 2D Fourier transform we can see that the $ss=6$ (or equivalently, FDTD) case only has a single mode present (i.e. a single peak in the Fourier space), centered at the Fourier component that is propagating at the speed of light. On the other hand, for $ss>6$ multiple modes are present, seen as multiple peaks in the Fourier space of various width and intensity, some of them propagating at speeds faster than the speed of light. The simulation was run until $n=300$ and the time direction was windowed with a Hann function before performing the Fourier transform. The displayed Fourier space plots are zoomed in, as the magnitude of the components not displayed was neglible. The Fourier transform convention used is $A_k = \sum_{m=0}^{N-1} a_m \exp(-2\pi i \ mk/N)$.

As a closing remark, it must be mentioned that the analysis here has been limited to a very special case of the hard source defined at a single point and the observations made are not general. Nevertheless, the problem considered turned out to be a suitable playground for the investigated method, as it was simple to implement and managed to highlight some shortcomings of our generalisation.

\section{Conclusions}
We have presented an attempt to generalise the FDTD method to a meshless setting, replacing its spatial derivatives by the appropriate RBF-FD approximations. Testing the method on the grid, we verified that this really is a generalisation - for the lowest possible stencil size we are able to reproduce the FDTD. We pointed out two major obstacles encountered in our generalisation - dispersion and stability. These are stencil-dependent properties and therefore especially important to understand in scattered node scenarios, where the shapes of stencils are less regular. In this context our paper serves as a basic, but necessary step towards developing a robust, adaptive scheme for solving electromagnetic problems meshlessly. A sensible continuation of this research is a focus on lowering the degree of dispersion or increasing the stability of the procedure, either by a careful selection of the stencils or an appropriate modification of our method.
\bibliographystyle{IEEEtran}
\bibliography{paper}

\end{document}